\documentclass[aps,pre,twocolumn,superscriptaddress,showpacs,amsmath,amssymb]{revtex4-1}

\usepackage{graphicx}

\begin{document}

\title{Rotational tumbling of {\it Escherichia coli} aggregates under shear}

\author{R. Portela}
\affiliation{UCIBIO@Requimte, Departamento de Ci\^encias da Vida, Faculdade de Ci\^encias e Tecnologia, Universidade Nova de Lisboa, 2829-516 Caparica, Portugal}

\author{P. Patr\'{\i}cio}
\affiliation{ISEL,
Rua Conselheiro Em\'{\i}dio Navarro 1, P-1959-007 Lisboa, Portugal}
\affiliation{CEDOC, Faculdade de Ci\^encias M\'edicas, Universidade Nova de Lisboa, 1169-056, Lisboa, Portugal}

\author{P. L. Almeida}
\affiliation{ISEL,
Rua Conselheiro Em\'{\i}dio Navarro 1, P-1959-007 Lisboa, Portugal}
\affiliation{CENIMAT/I3N, Faculdade Ci\^encias e Tecnologia, Universidade Nova de Lisboa, 2829-516 Caparica, Portugal}

\author{R. G. Sobral}
\affiliation{UCIBIO@Requimte, Departamento de Ci\^encias da Vida, Faculdade de Ci\^encias e Tecnologia, Universidade Nova de Lisboa, 2829-516 Caparica, Portugal}

\author{J. M. Franco}
\affiliation{Dpt. Ingenier\'{\i}a Qu\'{\i}mica, Facultad de Ciencias Experimentales, Universidad de Huelva, Spain}

\author{C. R. Leal}
\email{cleal@adf.isel.pt}
\affiliation{ISEL,
Rua Conselheiro Em\'{\i}dio Navarro 1, P-1959-007 Lisboa, Portugal}
\affiliation{CENIMAT/I3N, Faculdade Ci\^encias e Tecnologia, Universidade Nova de Lisboa, 2829-516 Caparica, Portugal}

\date{\today}

\begin{abstract}

Growing living cultures of {\it Escherichia coli} bacteria were investigated using real-time {\it in situ} rheology and rheo-imaging measurements.
In the early stages of growth ({\it lag phase}), and when subjected to a constant stationary shear, the viscosity slowly increases with the cell's population.
As the bacteria reach the {\it exponential phase} of growth, the viscosity increases rapidly, with sudden and temporary abrupt decreases and recoveries.
At a certain stage, corresponding grossly to the {\it late phase} of growth, when the population stabilizes, the viscosity also keeps its maximum  constant value, with drops and recoveries, for a long period of time. This complex rheological behaviour, which was observed to be shear strain dependent, is a consequence of two coupled effects: the cell density continuous increase and its changing interacting properties. Particular attention was given to the {\it late phase} of growth of {\it E. coli} populations under shear. Rheo-imaging measurements revealed, near the static plate, a rotational motion of {\it E. coli} aggregates, collectively tumbling and flowing in the shear direction.
To explain this behaviour, we introduce a simple theoretical model, in which the individual cells are transported by the flow, but remain rigidly attached to the other cells of the aggregate.

\end{abstract}

\maketitle

\section{Introduction}
\label{sec1}

The rheology of {\it Escherichia coli} suspensions and biofilms has been the focus of numerous research studies in recent years
\cite{elgeti2015physics,karimi2015interplay,mazza2016}.
These systems present a complex non-Newtonian rheological behaviour,
which depend not only on the ``passive'' physical properties of the bacteria, but also on their intrinsic ``activity''.
This ``activity'' may be in part associated with the individual swimming motion and propulsion apparatus of
each {\it E. coli} cell, beautifully described in \cite{gachelin2013non}, but also with the emergence of a collective behaviour,
usually related to cell density, but not exclusively.
The processes of bacterial chemotaxis and aggregation are driven by the availability of nutrients,
their own secretion of chemoattractants, which prevent the dispersion of the population \cite{saragosti2011directional},
the presence of oxygen \cite{lopez2015turning} and other external factors, as the
boundary confinement \cite{gachelin2013non}, the substrate surface roughness, or the external mechanical solicitation \cite{gachelin2013non,lopez2015turning}.

In some interesting situations, in diluted and in semi-diluted regimes, the ``active'' viscosity can be lower than the viscosity of the suspending fluid at low shear rates. The collective coordinated motion of the bacteria
seems to create a ``superfluid'' \cite{gachelin2014collective,lopez2015turning}.
In other situations, {\it E. coli} bacteria form biofilms --
self-organized, integrated communities composed of bacterial cells embedded in a matrix
of self-produced extracellular polymeric substances (EPS) --, which adhere to surfaces, and are capable
of withstanding chemical and mechanical stresses. Biofilms may evolve into complex configurations such as granules, ripples, or streamers,
which influence the mechanical response of the culture \cite{karimi2015interplay}.

Several experimental setups have been used to study the rheology of bacterial suspensions and biofilms \cite{karimi2015interplay}.
Some use (macroscopic) conventional rheometry, such as stress-strain, creep and small amplitude oscillatory shear tests, others use microfluidic devices, which allow access to the internal structure of the cell clusters.
Often, the performed studies are time-window confined and do not refer to a regular culture cell growth process.

In this work we study the rheology of \textit{E. coli} cultures during growth, {\it in situ}, when subjected to a stationary shear flow.
We have found three distinct rheological behaviours, corresponding to the three distinct phases of growth, the {\it lag, exponential} and {\it late} phases. In order to understand these different viscoelastic behaviours,  we used rheo-imaging to assess the cell's organization and patterning in each phase of growth, while the shear flow was being applied. Particular attention is addressed to the {\it late phase} of growth where the highest density of cells is attained and a surprising rotational tumbling motion of cell's aggregates, not previously reported in the literature, is observed. An attempt to quantitatively describe this behaviour is proposed, considering a theoretical model based on simple rigid body mechanics.

This article is organised as follows: in Section \ref{sec2} we describe the bacteria culture considered in this study, and the various experimental techniques used to characterise them: bacteria propagation, optical density measurements, rheology and rheo-imaging characterisation. In Section \ref{sec3} we analyse the rheological behaviour under steady-state shear flow. From the real-time rheo-imaging it is possible to obtain crucial information on bacteria self-aggregation patterns and intrinsic motion. In Section \ref{sec4} we describe the theoretical model proposed to describe the rotational tumbling motion observed in this {\it E. coli} culture at high cell density stage, and we present our conclusions in Section \ref{sec5}.

\section{Experimental / Methods}
\label{sec2}

\subsection{Bacteria strain and growth conditions}

{\it Escherichia coli} DH5$\alpha$ (Invitrogen,USA) was used. Cultures were grown at 37 $^\circ$C with aeration in LB (NZYtech,Portugal). Over-night grown cultures were used to re-inoculate fresh medium at an initial $\text{OD}_{620\text{nm}}$ of 0.005, for rheological characterisation.
To monitor bacterial growth, we measured the optical density (620 nm) at discrete time intervals, resorting to a spectrophotometer Ultrospec 2100 pro.
In parallel, we also determined the population colony forming units (cfus/ml), which provides an estimation of the viable cells, by plating serial dilutions of the bacterial cultures on LA (NZYtech,Portugal). The plates were incubated for 48 h at 37 $^\circ$C, and the colonies were counted. Growth of {\it E. coli} cultures in culture medium was monitored by measuring the optical density ($\text{OD}_{620\text{nm}}$) at discrete time intervals, in parallel with population's colony-forming units (cfus/ml).

\subsection{Rheology}

Rheological measurements were performed in a controlled stress rotational rheometer Bohlin Gemini HR$^\text{nano}$. A steel plate/plate geometry, with diameter 40 mm and 2000 $\mu$m gap (to ensure a good signal), was used for the measurements of the viscosity growth curve, at constant shear rate of 10 s$^{-1}$, at 37 $^\circ$C (to allow optimal bacterial growth). A solvent trap was used in all measurements to avoid evaporation.

\subsection{Rheo-Imaging}

Real-time  image  acquisition  was  performed  during  steady-state  shear  flow  measurements  in  a ThermoHaake  RheoScope  equipment,  which  combines  the  principles  of  a  conventional  controlled stress  rheometer  with  an  optical  microscope.  A  constant  shear  rate  of  10 s$^{-1}$ was  imposed  using  a  cone/plate  geometry  with 70 mm diameter and an angle of 1$^\circ$, at 37 $^\circ$C. The cone had a mirror surface and the plate a  cover glass, to allow optical microscopic observations (20x) during shear (schematic details are enclosed  in  Fig.  1),  at  an  intermediate  radius  plate  fixed  position.
A sequence of one photo image per min was extracted from the video. In these  tests, the  growth  of  an  \textit{E. coli} culture was followed by starting measurements already at the {\it exponential phase} (approximately at  an $\text{OD}_{620\text{nm}}$=2.5).
Video image acquisition was performed during 150 min. At the end of these measurements two videos with a higher frame rate acquisition (1 frame/s) were recorded, each with a duration of 3 min, under the same rheological conditions (supplementary material S1).

\section{Results and Discussion}
\label{sec3}

From the optical density and population's forming units measurements, it was possible to identify three different growth phases: {\it lag phase}, {\it exponential phase} and {\it late phase}. In the first stage of growth, corresponding to the {\it lag phase}, bacteria are adapting to the new environmental growth conditions. Secondly, the {\it exponential phase} of bacterial growth starts with the concentration of bacteria in the medium increasing exponentially. The last stage, {\it late phase}, occurs when the bacteria population starts to stabilize.

\begin{figure}[htp]
\begin{center}
\includegraphics[scale=0.4]{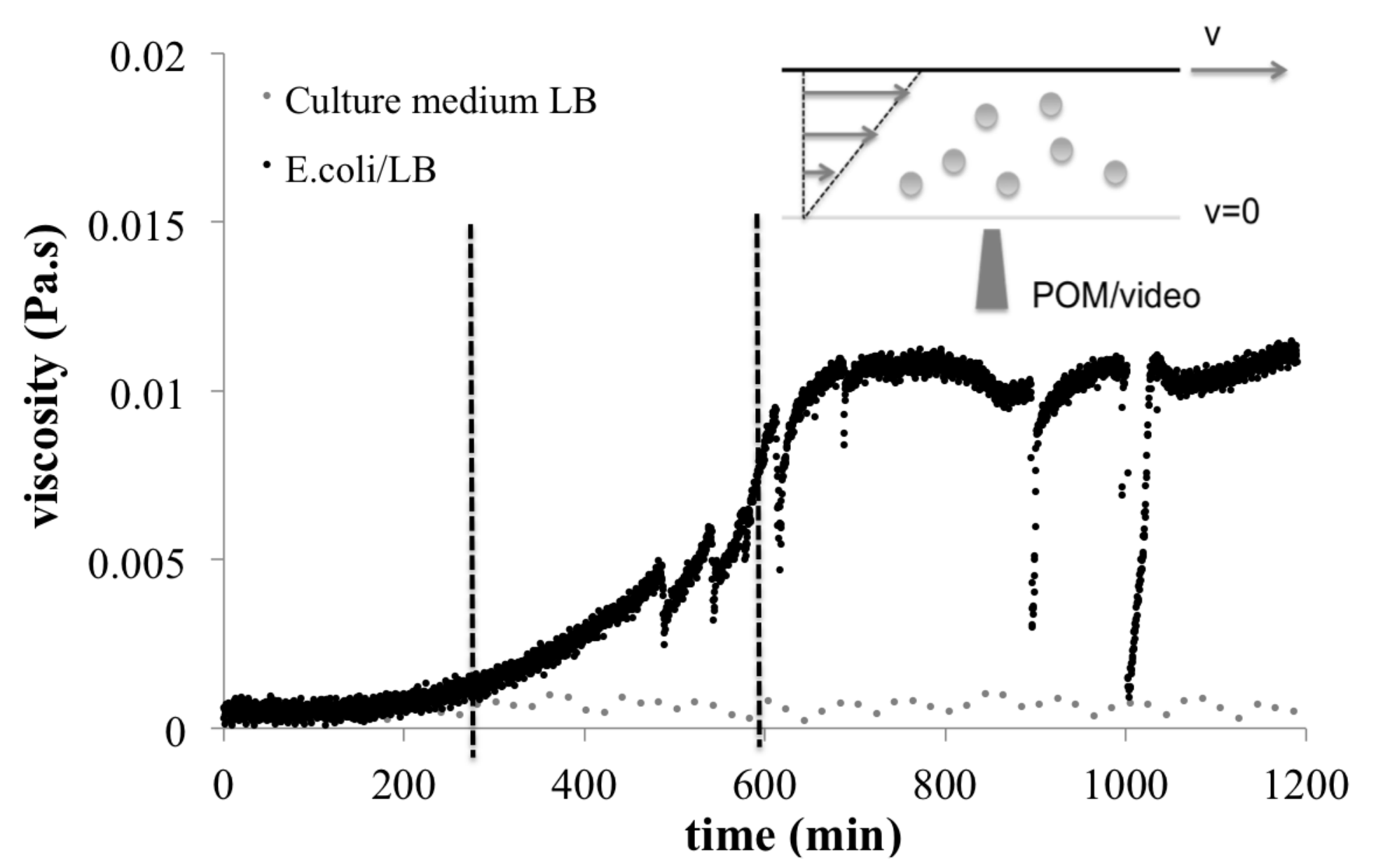}
\caption{ \textit{E. coli} cultures steady-state shear viscosity growth curve, $\eta(t)$ measured at a constant shear rate of  10 s$^{-1}$ (black) and culture medium (gray) (representative curves). Dashed lines separate distinct growth phases:  {\it lag phase}, {\it exponential phase} and {\it late phase}, determined by  $\text{OD}_{620\text{nm}}$ and population colony forming units (cfus/ml) measurements. All measurements performed at 37 $^\circ$C. Inset: schematic details on the image acquisition set-up at an intermediate radius / plate  position;}.
\label{fig1}
\end{center}
\end{figure}

The general viscosity growth curve, in Figure  \ref{fig1}, suggests a really complex and rich rheological behaviour, showing three distinct phases, consistent with the time intervals previously defined for the {\it lag}, {\it exponential} and {\it late phases}.

In the first 300 minutes of growth, corresponding to the {\it lag phase}, bacteria are adapting to the new environmental growth conditions, with a characteristic slow division rate. Accordingly, the viscosity shows a slow and constant increase with time. In the {\it exponential phase} (300 - 600 min), where the  $\text{OD}_{620\text{nm}}$  strongly increases, the viscosity presents a dramatic increase by a factor of 30 with respect to its initial value. The viscosity increase in this period is not monotonic, but exhibits several drops and recoveries.
At 600 min, it is known that the cfus/ml starts to stabilize, most probably due to nutrient depletion and accumulation of secondary metabolites which inhibit cell division. This corresponds to the beginning of the {\it late phase}. At approximately this time point, the viscosity increase slows down,
reaching an intermittent plateau of maximum viscosity,  with several drops and recoveries.

Comparable results were obtained for {\it S. aureus} - strain COL culture \cite{portela2013,patricio2014living}:
In this case, we also observed in the viscosity growth curve three distinct regimes, corresponding to the {\it lag}, {\it exponential} and {\it late phases}.
The {\it lag phase} with a slow viscosity increase, the {\it exponential phase} with a dramatic viscosity increase, exhibiting also several sudden drops and recoveries,
and a {\it late phase}, in which the viscosity decreases. However, in contrast with the \textit{E. coli} results, the viscosity decrease of \textit{S. aureus} at
the {\it late phase} was abrupt, recovering almost its initial value in a very short period (approximately 100 minutes).
The {\it S. aureus} bacteria have a spherical shape with 1 $\mu$m diameter, they are not self-motile, and they form small clusters of  5-15 cells.
As the cell density increases, the bacteria aggregates start to establish new contacts and form frequently a web of cellular structures \cite{franco2015cell}.
This may explain the initial viscosity increase observed in the {\it lag} and {\it exponential phases}. The viscosity drops and recoveries
at the {\it exponential phase} may also be associated with the formation and disruption of the web cellular structures during shear.
It is known that in the {\it late phase} the bacteria diminish substantially the production of adhesins \cite{voyich2005insights}.
Without being able to adhere, the bacteria sediment in the static plate (as observed with rheo-imaging \cite{franco2015cell}),
justifying the abrupt viscosity decrease at this phase of growth.

In the present case of an {\it E. coli} culture, the viscosity increase observed in the {\it lag} and {\it exponential phases}
seems also naturally related to the cell density increase. In the {\it late phase}, however,
the viscosity remains approximately constant (although with sudden drops and recoveries) and we do not observe bacteria sedimentation.
Instead, rheo-imaging measurements of growing {\it E. coli} populations under shear, in the {\it late phase}, revealed a collective rotational motion, associated with the translational motion in the shear direction, near the static plate (this collective motion was not identified for the small {\it S. aureus} clusters).
Different size aggregates were observed at this stage of growth, where most of the bacteria are moving together and bacteria deposition rarely occurred. A possible understanding for the formation of this aggregates follows. As the individual cells contact with the plate, they adhere, creating a thin cell layer. These thin cell planar aggregates frequently detach from the plate, and start to move collectively, sometimes curling themselves in a cylindrical hollow shape, rotating around the direction that is perpendicular to the plane of the flow (the vorticity direction), and moving with the fluid with a constant linear velocity. The aggregates' movement and its interaction
with the transient thin layers may justify the viscosity drops and recoveries during the {\it exponential} and {\it late phases}.
As we do not observe cell deposition, the bacteria density in the bulk remains approximately constant at the  {\it late phase}, justifying the viscosity plateau.

\begin{figure*}
\begin{center}
\includegraphics[scale=1]{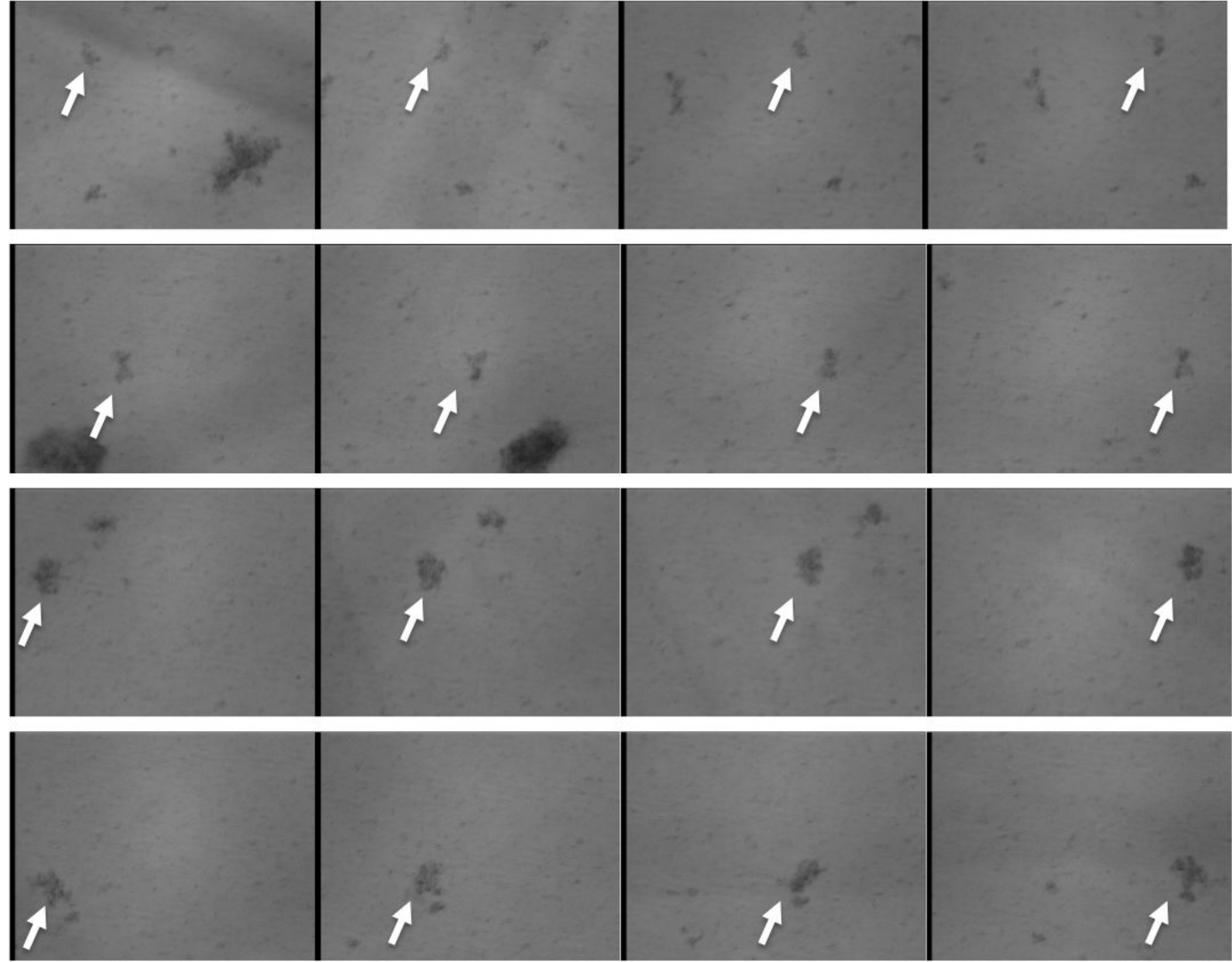}
\caption{Each row corresponds to a 4-image sequence (1 image/s) extracted from Rheo-imaging measurements of an \textit{E. coli} culture,
in order to illustrate the aggregates rotational motion (the rotating aggregates are indicated by the white arrows and a full rotation may be seen for each sequence);
the images were collected during steady-state shear, at  a  constant  shear  rate  of  10 s$^{-1}$,
in the {\it late phase} of growth; each image has 225 $\mu$m of width (640x480 pixels); measurements were performed at 37$^\circ$C.
A full video may be seen in supplementary material S1.}
\label{fig2}
\end{center}
\end{figure*}

From the recorded videos, it was possible to characterise quantitatively the rotational and translational motion of the aggregates over a convenient statistical sample. Four examples are included in Figure \ref{fig2} and the average value of the angular velocity of the aggregates was estimated to be $2,2 \pm 0,7$ rad/s. The error associated to this estimation is mainly due to the acquisition frame rate, which is 1 frame/s.

In spite of the well-known {\it E. coli} intrinsic motility,
we do not observe an explicit individual motion of the cells within each aggregate. Instead, they seem to keep
their relative positions, suggesting they are connected by adhesive factors \cite{karimi2015interplay}.
The role of the {\it E. coli} intrinsic motility is surely an important ingredient to understand the rheological behavior of the colonies.
In particular, we do not observe a significant {\it E. coli} sedimentation, during the {\it late phase} of growth,
as occurred in {\it S. aureus}, a non-motile bacteria.
The relation between motility and sedimentation should be further investigated in future studies.

\section{Theoretical model}
\label{sec4}

Rheo-imaging measurements showed us unequivocally the existence of aggregates with a rotational tumbling motion, not previously reported
in the literature. The observed motion resembles the well-known tumbling motion of an ellipsoidal particle immersed in a viscous fluid,
described by Jeffery \cite{jeffery1922motion}. In our experiments, however, the rotational motion does not correspond to a single body,
but to a whole aggregate of cells, which seem strongly interconnected.
In order to physically describe this motion, and in a first approximation,
we modelled the cells as individual particles with the same mass, connected by rigid links or filaments,
and thus forming a global rigid body. Due to the small size of the cells, inertial effects are neglected.
Moreover, despite the large number of cells and their links, in this approximation the flow stays unperturbed.
Under a simple shear, the fluid velocity field is given by:
\begin{equation}
\vec u(\vec r)=\dot\gamma y \hat i
\end{equation}
where $\dot\gamma$ is the constant shear rate.
The position and velocity of each cell are
\begin{equation}
\vec r_i=r_{ix}\hat i+r_{iy}\hat j+r_{iz}\hat k;\;\;\;\;\;\vec v_i=v_{ix}\hat i+v_{iy}\hat j+v_{iz}\hat k.
\end{equation}
and $i=1,...,N$. The drag force on each cell may be written:
\begin{equation}
\vec F_i=b(\vec u(\vec r_i)-\vec v_i)=b(\dot\gamma r_{iy}\hat i-\vec v_i)
\end{equation}
where $b$ is the drag coefficient, which we take to be approximately constant, independent of the cell, and consequently, of its individual orientation.

In this simplified model, the cell's aggregate behaves as a rigid body. Neglecting inertial terms, the equations of motion are
\begin{eqnarray}
&\sum\limits_{i=1}^N\vec F_i=0\\
&\sum\limits_{i=1}^N\vec r_{Gi}\times\vec F_i=0
\end{eqnarray}
where $\vec r_{Gi}=\vec r_{i}-\vec r_{G}$, and $\vec r_{G}$ is the aggregate's center of mass.

The first equation  gives the motion of the center of mass, which has a constant velocity:
\begin{eqnarray}
\vec v_G=\dot\gamma r_{Gy}\hat i
\end{eqnarray}
The second equation gives the rotational motion of the aggregate, with angular velocity $\vec \omega$:
\begin{eqnarray}
\sum\limits_{i=1}^N\vec r_{Gi}\times(\dot\gamma r_{iy}\hat i-\vec \omega\times \vec r_{Gi})=0
\end{eqnarray}
If the vorticity direction coincides with one of the principal directions of the moment of inertia,
\begin{eqnarray}
\sum\limits_{i=1}^N\vec r_{Gi}\times r_{iy}\hat i=-\sum\limits_{i=1}^Nr_{iy}^2\hat k
\end{eqnarray}
and the aggregate rotates only in the plane of the flow, $\vec\omega=\dot\theta\hat k$. The second equation of motion becomes:
\begin{eqnarray}
\dot\gamma I_x+\dot\theta I=0
\end{eqnarray}
where the second order moments are defined by
\begin{eqnarray}
I=I_x+I_y=\sum\limits_{i=1}^N r^2_{Giy}+\sum\limits_{i=1}^N r^2_{Gix}
\end{eqnarray}
The second order moments define a 2D tensor which has principal moments $I_1$ and $I_2$,
associated with the principal directions $x'$ and $y'$, and related to $x-y$ through the rotation $\theta$.
Thus, we may write
\begin{eqnarray}
&I_x=I_1\cos^2\theta+I_2\sin^2\theta\\
&I=I_1+I_2.
\end{eqnarray}
The principal moments of the rigid aggregate are constant throughout the motion.
Its use explicits the dependence of the rotational angle $\theta$. The second equation of motion may now be written:
\begin{eqnarray}
\dot\gamma \left(I_1\cos^2\theta+I_2\sin^2\theta\right)+\dot\theta I =0,
\label{eq Jeffery}
\end{eqnarray}
Interestingly, this equation corresponds to the classical Jeffery equation \cite{jeffery1922motion}
for the motion of an ellipsoidal particle immersed in a viscous fluid, obtained by first solving Stoke's equations for the fluid around the particle
(a non trivial calculation).
In Jeffery equation, the second moments  are replaced by equivalent expressions involving the lengths of the ellipsoidal axes.
Instead, our simplified model does not specify the geometry of the rigid aggregate, but the fluid stays unperturbed.

If we define the relative difference between the principal second order moments $d=(I_2-I_1)/(I_1+I_2)$ (so $|d| \le 1$),
Eq. (\ref{eq Jeffery}) may be simplified to
\begin{eqnarray}
\frac{d\theta}{dt}=-\frac{\dot\gamma}{2}\left(1-d\cos 2\theta\right)
\label{eq Jeffery 2}
\end{eqnarray}
This equation may be integrated (it was firstly found by Jeffery).
If we rewrite Eq. \ref{eq Jeffery 2},
\begin{eqnarray}
\dot\gamma dt=\frac{-2d\theta}{1-d\cos 2\theta}\Leftrightarrow \dot\gamma t=\int_{\theta_0}^{\theta(t)}\frac{-2d\theta}{1-d\cos 2\theta}
\end{eqnarray}
Omitting a constant, we have the solution:
\begin{eqnarray}
\tan \frac{\sqrt{1-d^2}}{2}\dot\gamma t=\frac{1+d}{1-d}\tan\theta
\label{Jeffery_solution}
\end{eqnarray}
We have periodic solutions, with a periodicity that may be obtained directly from the previous integral expression:
\begin{eqnarray}
\dot \gamma T=\int_{\theta_0}^{\theta_0-2\pi}\frac{-2d\theta}{1-d\cos 2\theta}=\frac{4\pi}{\sqrt{1-d^2}}
\end{eqnarray}
Thus, the absolute value of the mean angular velocity of the aggregate (rotating clockwise) is
\begin{eqnarray}
\dot\theta_m=\frac{2\pi}{T}=\frac{\dot\gamma}{2}\sqrt{1-d^2}
\end{eqnarray}
and should be always smaller than half the shear rate.

In our experimental results, the distribution of bacteria in the projected plane of the rotation is approximately circular,
meaning $I_1\approx I_2$ and  $d$ small. In this limit, we have the explicit approximate solution:
\begin{eqnarray}
\dot\theta\approx -\frac{\dot\gamma}{2}(1-d\cos \dot\gamma t)
\end{eqnarray}
If we had a very asymmetric collection of bacteria, with $d\approx 1$,
$\dot\theta\approx 0$ for $\theta\approx n\pi$. In this case, the aggregate
would be oriented most of the time along the axis with smaller second order moments.
Eventually, the aggregate would acquire larger and larger absolute angular velocities, reaching
$\dot\theta=-\dot\gamma$ for $\theta=\pi/2+n\pi$, and slowing down again. The motion is periodic, but with larger and larger periodicities as
$d\to 1$.

In our experimental results, we have a small number of frames per aggregate's complete revolution.
It is thus difficult to have a precise idea about its second order moments.
The measured mean angular velocity presents a large error ($\dot\theta_m\approx 2.2\pm0.7$ rad/s),
and is roughly $1/4$ of the shear rate ($\dot\gamma\approx10\pm 0.01$ 1/s).
This would correspond to a fairly asymmetric aggregate
(in the rotational plane, which is perpendicular to the plane of the images, and leading to a large value of $d$), not suggested by the pictures (see Fig. \ref{fig2}).
We are currently planning new optical images, in more precise experiments.
From the theoretical point of view, we could think of more developed models, using for instance chemioattractive interactions
({\it a la} Keller-Segel type \cite{keller1971model, saragosti2010mathematical}),
or active brownian forces, expressing the cell's collective interaction, added to the equation of motion of each individual cell
\cite{saragosti2011directional}. Nevertheless, from our images, we have the impression that the motion of the aggregates resembles
the motion of a rigid body. Without other accurate measurements, we think this model seems to retain the fundamental
aspects of our results.

\section{Conclusions}
\label{sec5}

In this study, real-time and in situ rheo-imaging rheology was applied to the animal commensal bacteria
{\it E. coli} during cell growth. As the density of bacteria in the medium increases, cells may rearrange themselves in different aggregates, capable of strongly influencing their environment, and leading to three different physical rheological responses,
corresponding to the three distinct phases of growth, the {\it lag, exponential} and {\it late} phases.

In the {\it lag phase}, bacteria are adapting to the new environmental growth conditions, with a characteristic slow division rate. Accordingly, the viscosity shows a slow and constant increase with time. In the {\it exponential phase} the viscosity presents a dramatic increase, but exhibits several drops and recoveries.
In the {\it late phase} of growth, the viscosity increase slows down,
reaching an intermittent plateau of maximum viscosity, with several drops and recoveries.
In this phase, the highest bacteria density is attained -- bacteria still grow and divide, but at a lower rate.
Big and irregular bacteria aggregates are observed, which keep moving in suspension. No significant sedimentation is observed.
The aggregates present translational motion in the shear flow direction, and rotational motion in the vorticity direction.
The aggregates become larger in time, due to the incorporation of smaller aggregates. Due to the rotational motion,
the aggregates become elongated along the rotational axis. Apparently, the size of the aggregates does not influence the rotational motion, since almost all aggregates rotate with the same angular velocity, which is related with the applied shear rate.
In spite of the well-known {\it E. coli} intrinsic motility, the proposed model based on rigid-body motion seems to offer a sufficient physical description to the rotational motion of the {\it E. coli} aggregates.

\section*{Acknowledgements}

This work was also supported by FEDER through the COMPETE 2020, and FCT projects UID/CTM/50025/2013,
PTDC/FIS-NAN/ 0117/2014 (awarded to PLA) and ESCMID grant 2015 (awarded to RGS)
and by the ``Unidade de Ci\^encias Biomoleculares Aplicadas- UCIBIO" which is financed by national funds from FCT/MEC
(UID/Multi/ 04378/2013) and co-financed by the ERDF under the PT2020 Partnership Agreement (POCI-01-0145-FEDER-007728 ).

\bibliography{bacteria}

\end{document}